\documentclass[useAMS,usenatbib]{mn2e}

\usepackage{graphicx}
\title{Self-similar galaxy dynamics below the de Sitter scale of acceleration}

\author[Maurice, H.P.M. van Putten]
{Maurice H.P.M. van Putten$^1$ \\ 
$^{1}$Physics and Astronomy, Sejong University, 98 Gunja-Dong Gwangin-gu, Seoul 143-747, Korea}

\begin{document}

\date{}

\pagerange{\pageref{firstpage}--\pageref{lastpage}} \pubyear{2002}

\maketitle

\label{firstpage}

\begin{abstract}
Radial accelerations $\alpha$ in galaxy dynamics are now observed over an extended range in redshift that includes 
model calculations on galactic distributions of cold dark matter (CDM) in $\Lambda$CDM. In a compilation of data 
of the {\em Spitzer} Photometry and Accurate Rotation Curves (SPARC) catalogue, the recent sample of Genzel et 
al.(2017) and the McMaster Unbiased Galaxy Simulations 2, we report on effective self-similarity in the variable 
$\zeta = a_N/a_{dS}$, given by the Newtonian acceleration $a_N$ based on baryonic matter content over the de 
Sitter scale of acceleration $a_{dS}=cH$, where $c$ is the velocity of light and $H$ is the Hubble parameter. 
SPARC, MUG2 and theory satisfy ${a_N}/{\alpha} \simeq 2.1\,\zeta^\frac{1}{2}$ $(\zeta <<1)$. 
At $\zeta=1$ in transition to Newtonian gravity ($\zeta>>1$), however, there is a $6\sigma$ gap between SPARC 
and MUGS2. This poses a novel challenge to CDM in $\Lambda$CDM against the apparent $C^0$ galaxy dynamics 
observed in SPARC. We attribute the latter to reduced inertia below 
the de Sitter scale of acceleration $(\zeta < 1)$, based on a causality constraint imposed by the cosmological 
horizon ${\cal H}$. 
\end{abstract}


\section{Introduction} 

Advances in high resolution spectroscopy of galaxy rotation curves across a range of redshifts give
a detailed view on radial accelerations over an extended range in radius $r$ and redshift $z$
up to about two \citep{fam12,lel16,mcg16,gen17}.
In $\Lambda$CDM, these observations suggest a diminishing of cold dark matter content with $z$, as 
observed accelerations $\alpha$ increasingly match the Newtonian acceleration 
\begin{eqnarray}
a_N=\frac{GM_b}{r^2}
\label{EQN_aN}
\end{eqnarray} 
by baryonic matter content $M_b$ within $r$, where $G$ is Newton's constant.
These results provide important benchmarks for galaxy models in $\Lambda$CDM from high resolution smoothed particle hydrodynamics
simulations of galaxy formation. A recent comparison of the McMaster Unbiased Galaxy Simulations 2 (MUGS2) sample of galaxy models, for instance, suggests excellent agreement with the ``missing mass" in galaxy rotation curves from the {\em Spitzer} Photometry 
and Accurate Rotation Curves (SPARC) \citep{kel17}. Here, we revisit this claim focused on the transition regime of gravitational
acceleration consistent with Newton's theory based on baryonic matter and weak gravitation, marked by anomalous dynamics
commonly attributed to dark matter or a modification of Newtonian gravitation \citep{fam12}.

In a model-independent approach, redshift dependence in galaxy dynamics shows evolution with background 
cosmology described by the Hubble parameter $H=H(z)$, carrying a de Sitter scale of acceleration
\begin{eqnarray}
a_{dS}=cH,
\label{EQN_adS}
\end{eqnarray}
where $c$ is the velocity of light and $H$ is the Hubble parameter (Fig. 1). For a galaxy such as the Milky Way,  
 $a_N=a_{dS}$ corresponds to a distance \citep{van16}
\begin{eqnarray}
r_t = \sqrt{R_gR_H} = 4.6\,\mbox{kpc} \,M_{11}^{1/2},
\label{EQN_rt}
\end{eqnarray}
where $R_H=c/H$ is the Hubble radius in a three-flat Friedmann-Robertson-Walker universe and $R_g=GM/c^2$ is
the gravitational radius of a galaxy of mass $M=M_{11}10^{11}M_\odot$. In quantum cosmology, $a_{dS}$ represents
the surface gravity of the cosmological horizon at Hubble radius $R_H=c/H$ in de Sitter space \citep{gib77}. 
Based on dimensional analysis, this suggests evolution in galaxy dynamics in 
\begin{eqnarray}
\zeta = \frac{a_{N}}{a_{dS}},
\label{EQN_xi}
\end{eqnarray}
where $\zeta=1$ corresponds to a collusion of Rindler and cosmological horizon \citep{van17b}. 

{We here consider data on galaxy dynamics as a function of $\zeta$, of galaxy rotation curves
of observed galaxies and numerical galaxy models in $\Lambda$CDM side-by-side (\S2). This compilation 
highlights {\em self-similar behavior} in galaxy dynamics in $\zeta$ - by which data over different redshifts
coalescence - and a transition across $\zeta=1$ to 
weak gravitation ($\zeta<1)$ from normal, Newtonian gravitation ($\zeta>1)$, where 
observed and modeled galaxy dynamics differ.} These observations are interpreted in \S3.
This study is restricted to late-time cosmology with redshifts $z$ up to about two, over
which range the Hubble parameter varies by a factor up to about three (Fig. 1).
In \S4, we give our conclusions and outlook on future observations.

\section{Self-similar galaxy dynamics}

Galaxy rotation data considered here are taken from SPARC \citep{mcg16,lel16}, MUGS2 \citep{kel17} and
\cite{gen17}.

SPARC provides a sample of rotation curves observed from 175 nearby mostly late Hubble type 
galaxies observed by spectroscopy and photometry in Hi/H$\alpha$, covering a broad range in 
luminosity $(10^{7-12}L_\odot)$, radii ($0.3-15$\,kpc), effective surface brightness 
($5-5000L_\odot$pc$^{-2}$) and rotation velocities ($20-300$ km\,s$^{-2}$) consistent with a 
stellar mass-to-light ratio $0.5M_\odot/L_\odot$.

MUGS2 provides a sample of 18 galaxy models with halo masses $3.7\times 10^{11}-2.2\times10^{12}M_\odot$
and disk masses $1.8\times10^{10}-2.7\times10^{11}M_\odot$ in a $\Lambda$CDM cosmology 
from high resolution smoothed particle hydrodynamics simulations with radiative cooling, star-formation and 
feedback from supernovae \citep{wad04,vol05,she10,kel14}. Excluding galaxies that experience 
appreciable tidal interactions and limited to galaxies modeled by at least 100 star particles, it is 
extended to a total of 32 galaxies at $z=0$ \citep{kel17}.

\cite{gen17} provides a sample of six rotation curves of galaxies at intermediate redshifts
$z\,\epsilon\,\{ 0.854, 1.5, 1.613, 2.196, 2.242, 2.383\}$ with respective baryonic masses
$M_{b,11} = \{ 1.7, 2.3, 1.0, 1.7, 1.7, 2.1\}$ and $\Lambda$CDM Hubble parameters
$H(z)/H_0=\{ 1.599, 2.288, 2.425, 3.190, 3.253, 3.454\}$ featuring rotation velocities
$V_c = \{276, 310, 257, 301, 364, 299\}$\,km/s at radii $R_{1/2} = \{ 7.3, 7.4, 4.9, 5.5, 3.3, 6\}$\,kpc.
Following (\ref{EQN_xi}), their $\zeta$ values cluster about $\zeta =1$ \citep{van17b},
\begin{eqnarray}
\zeta =\{ 0.2942,  0.3100,  0.3162, 0.3378, 0.4034,  0.8521\}.
\label{EQN_xiG}
\end{eqnarray}

\begin{figure}
\begin{center}
  \includegraphics[scale=0.49]{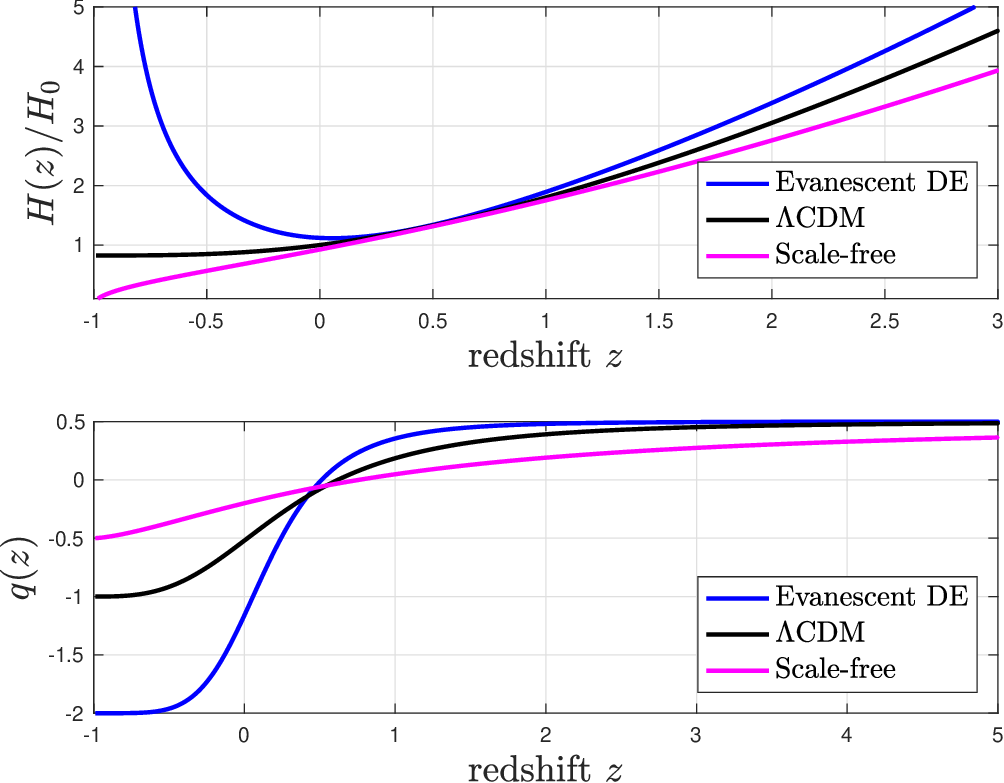}
\end{center}
\caption{Evolution of the Hubble parameter $H$ in cosmologies with different models of dark energy and
the associated deceleration parameter $q(z)=-1+(1+z)H^{-1}(z)H^\prime(z)$ as a function of redshift $z$. 
Next to $\Lambda$CDM are shown evolution by evanescent dark energy ($\Lambda=\omega_0^2$ set
by the eigenfrequency $\omega_0^2$ of the cosmological horizon \citep{van17b}) and a scale-free
cosmology \citep{mae17,jes17}. These models have similar evolution in the past ($z>0$), showing
an increase in $H$ by a factor of about two (three) at $z=1$ ($z=2$), while showing dramatically different behavior in the future $(-1<z<0)$.}
\label{fig1}
\end{figure}

\begin{figure}
\begin{center}
\includegraphics[scale=0.49]{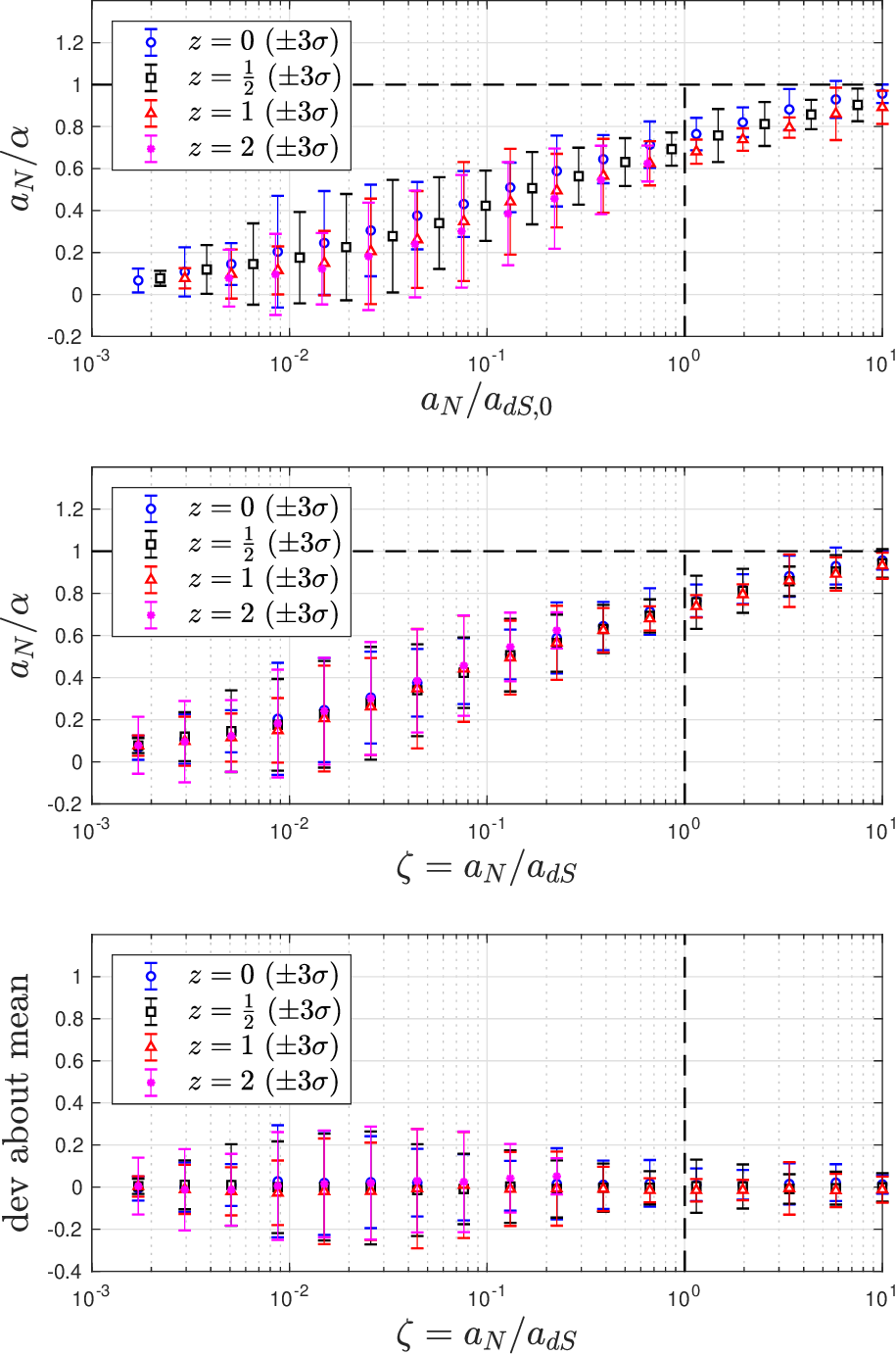}
\end{center}
\caption{
Compilation of radial accelerations in MUGS2 data \citep{kel17} in $\Lambda$CDM numerical galaxy models plotting rotation curve data as $a_N/\alpha$ versus $a_N/a_{dS,0}$ (top panel) and versus the 
similarity variable $\zeta = a_N/a_{dS}$ (middle panel), covering weak gravity ($a_N/a_{dS}<1$) and 
Newtonian gravity $a_N/a_{dS}\ge1$ of baryonic matter content ({\em black dashed}). The results {show remarkable self-similarity} in $\zeta$, in galaxy evolution tracing background cosmology with Hubble 
parameter increasing by a factor of a few over $0\le z \le 2$. For $z=2$, data are limited to
$\zeta \le 0.2254$. For each $z$, deviations from the mean of rotation curve data (over $z=0,\frac{1}{2},1$)
are much smaller than statistical errors in this average (bottom panel).}
\label{fig2}
\end{figure}

\begin{figure}
\begin{center}
\includegraphics[scale=0.49]{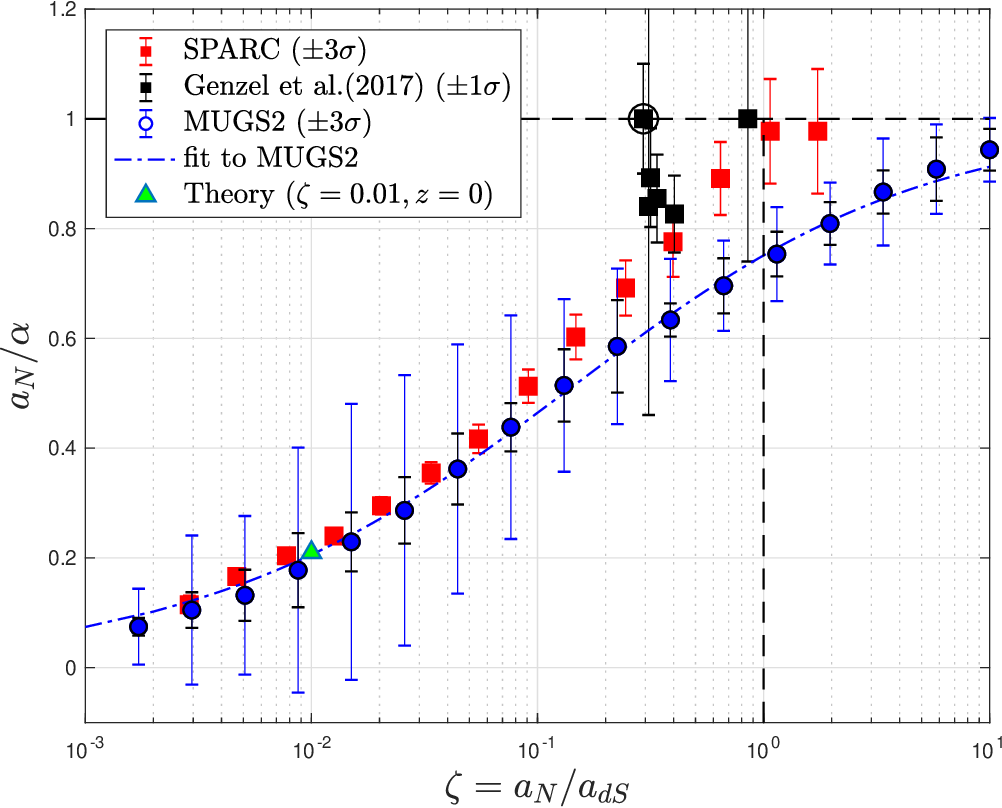}
\end{center}
\caption{
Radial accelerations in spiral galaxies from SPARC ({\em red}), Genzel et al.(2017) ({\em black})
and MUGS2 ({\em blue}: $a_N/\alpha$ versus $\zeta=a_N/a_{dS}$ covering weak gravity ($a_N/a_{dS}<1$) and Newtonian gravity $a_N/a_{dS}\ge1$ of baryonic matter content ({\em black dashed}). 
Error bars are $3\sigma$ for data of SPARC and MUGS2 (uncertainties from scatter ({\em blue}) and averaging over
redshift ({\em black})) and $1\sigma$ for \citep{gen17}. 
SPARC covers $z=0$, MUGS2 $z_k=0,\frac{1}{2},1$ and $z=2$ ($\zeta \le 0.2254$), and Genzel et al. galaxies have 
$\zeta$ clustered about $\zeta =1 $. %=\{ 0.2942,  0.3100,  0.3162, 0.3378, 0.4034,  0.8521\}$. 
For the latter, highlighted is the galaxy cZ 4006690 ({\em black circle}), which may have systematic errors in the observed strong asymmetry of its rotation curve. Observations and simulations {show remarkable self-similarity} in $\zeta$, in galaxy evolution tracing background cosmology over redshifts up to a few. SPARC, MUGS2 and theory (Eq. 6) agree asymptotically $\zeta<<1$. At $\zeta=1$, SPARC appears to show $C^0$ galaxy dynamics (continuous with discontinuous derivative) \citep{van17a,van17b}, while MUGS2 provides a smooth interpolation with a 6$\sigma$ discrepancy at $\zeta=1$.
}
\label{fig3}
\end{figure}

Fig. 2 shows a compilation of MUGS2 rotation curve data {plotted as a function of $\zeta$}. Over $0\le 0 \le 2$, $H(z)$ varies by a factor of about three, implying variations of order unity in dimensional quantities such as $r_t$. For MUGS2, averaging of rotation curve data
$(\zeta,a_N/\alpha)_{z_k}$ over different redshifts leaves a dispersion much smaller than scatter in the data. 

Fig. 3 shows a compilation of the three galaxy samples of SPARC, \cite{gen17} and MUGS2 combined. While there is excellent agreement between SPARC and MUGS2 in the weak gravity limit $\zeta <<1$, there appears to be an appreciable {\em gap} about $\zeta = 1$ in transition to the Newtonian limit $\zeta >>1$. As a function of $\zeta$, the relatively high redshift data from \cite{gen17} agree within uncertainties with SPARC except for the outlier cZ 4006690.

Plotted as a function of $\zeta$, the aforementioned ``missing mass" in galaxy rotation curves appears to be {\em self-similar} over an extended range of redshift, absorbed in normalization by $a_{dS}$ giving a {\em reduction in 
independent variables by one}.

In the outskirts of galaxies, rotation curves satisfy \cite{mil83}'s law, $\alpha=\sqrt{a_0a_N}$ with
\citep{van17b} 
\begin{eqnarray}
a_0 = \frac{\omega_0}{2\pi},
\label{EQN_a0}
\end{eqnarray}
based on the fundamental frequency $\omega_0 = \sqrt{1-q}\,a_{dS}$ of the Hubble horizon, 
where $q=q(z)$, $q(z)=-1+(1+z)H^{-1}(z)H^\prime(z)$
is the deceleration parameter. In the asymptotic regime $\zeta<<1$, therefore, we have
\begin{eqnarray}
\frac{a_N}{\alpha} = (2\pi)^\frac{1}{2}(1-q)^{-\frac{1}{4}}\zeta^{1/2}\simeq 2.1\,\zeta^\frac{1}{2},
\end{eqnarray}
where the right hand side refers to $z\simeq0$. Deviations from self-similarity by $\pm20\%$ in 
$(1-q)^\frac{1}{4}$ as $q$ varies over $-1<q<0.5$ in late-time cosmology are too small to be resolved 
in the present data.

 \section{A $6\sigma$ gap at $\zeta=1$} %Causality constraint on inertia}

In transition from Newtonian gravity (\ref{EQN_aN}) ($\zeta>>1$) to weak gravity ($\zeta << 1$),  Fig. 3 
shows an onset to the latter which is smooth in MUGS2 in contrast to what appears to be
$C^0$ galaxy dynamics - continuous with discontinuous derivatives - in SPARC \citep{van17b}. 
(Uncertainties in the data do not resolve whether the transition is truly $C^0$ or nearly so.) 
Smoothness in MUGS2 is expected and inherent to $N$-body simulations by diffusion due to small 
angle gravitational scattering and gas dynamics.
The noticeable gap between MUGS2 and SPARC at $\zeta=1$ hereby might be 
characteristic for galaxy models in $\Lambda$CDM, not limited to MUGS2.

It is perhaps paradoxical, that $\zeta$ is a similarity variable familiar from the theory of linear diffusion, yet
the apparent $C^0$ onset to weak gravity in SPARC runs counter to the same. We attribute this result to
a break in Newton's second law - assuming a constant inertia at arbitrarily small accelerations - 
on a cosmological background with finite Hubble radius $R_H$, equivalently
attributed to thermodynamic properties of the associated cosmological horizon ${\cal H}$. 

According to the equivalence principle of general relativity, inertia can be identified with inertial mass-energy \citep{van17b}
\begin{eqnarray}
U = mc^2
\label{EQN_U}
\end{eqnarray} 
given by the gravitational binding energy in the gravitational field over the distance
\begin{eqnarray}
\xi=\frac{c^2}{\alpha}
\label{EQN_xi}
\end{eqnarray}
to the Rindler horizon $h$ at a given acceleration $\alpha$. Here, $U$ obtains
by integrating the inertial force $F=m\alpha$ over a distance $\xi$.

According to quantum field theory, the vacuum seen by a Rindler observer is 
described by a finite temperature diffusion constant \citep[reviewed by][]{son07} 
\begin{eqnarray}
D = \frac{\hbar c^2}{2\pi k_BT}
\label{EQN_D}
\end{eqnarray}
where $\hbar$ is Planck's constant and $k_B$ is the Boltzmann constant. With 
$D=\xi c$, (\ref{EQN_D}) is the thermodynamic interpretation of Rindler's relation
(\ref{EQN_xi}) at the Unruh temperature $T=T_U$ \citep{unr76},
\begin{eqnarray}
k_BT_U = \frac{\hbar \alpha}{2\pi c}.
\label{EQN_TU}
\end{eqnarray}
Identifying $T_U$ with the temperature of $h$, $U$ derives from the entanglement entropy 
$I_1 = 2\pi \Delta \varphi_C$, where $\Delta \varphi_C$ is the distance 
$\xi$ expressed in Compton phase \citep{van15}, giving
\begin{eqnarray}
U = \int_0^\xi T_U dI_1,
\label{EQN_Uh}
\end{eqnarray}

In (\ref{EQN_Uh}), $h$ is an apparent horizon surface.
Apparent horizon surfaces are familiar concept in numerical relativity signaling black hole 
formation \citep{pen65,bre88,coo00,yor89,wal91,coo92,tho07}. In a three-flat Friedmann-Robertson-Walker
universe, the cosmological horizon ${\cal H}$ provides an apparent horizon in the background, whose
Hubble radius $R_H$ puts a bound on $\xi$.
As $h$ formally drops beyond ${\cal H}$ ($\xi > R_H$), $U$ in (\ref{EQN_U}) drops below its 
Newtonian value $m=m_0$, since integration over the gravitational field is cut-off at $R_H$ by
${\cal H}$ as a causal boundary on $D$ in (\ref{EQN_D}).
At a given $a_N$, the observed acceleration 
\begin{eqnarray}
\alpha = \left( \frac{m_0}{m}\right) a_N
\label{EQN_alpha}
\end{eqnarray}
experiences a $C^0$ transition across $\zeta =1$ \citep{van17a,van17b}. 
The apparent $C^0$ galaxy dynamics in the SPARC data can hereby be 
attributed to causality imposed on inertial mass-energy by ${\cal H}$, leading to
a $6\sigma$ gap at $\zeta = 1$ between it and MUGS2.

\section{Conclusions}

An effective self-similarity $\zeta$ in galaxy dynamics enables a comprehensive confrontation between galaxy rotation curves
from observations and simulations over an extended range of redshifts, here shown in Fig. 3 for SPARC, MUGS2 and 
Genzel et al. galaxies covering redshifts up to about two. In weak gravity $(\zeta<<1)$, SPARC, MUGS2 and theory agree.
At $\zeta=1$, however, there is a $6\sigma$ gap between SPARC and MUGS2, where the first appears to show
$C^0$ galaxy dynamics while the second gives a smooth transition between $\zeta <<1$ and the Newtonian regime $\zeta >> 1$. 
In Fig. 3, the latter is emphasized by a simple fitting function
\begin{eqnarray}
\frac{a_N}{\alpha} = \left(  \frac{1}{2} + \sqrt{ \frac{1}{4} + \frac{1}{x} + \frac{1}{\sqrt{x}} } \right)^{-1}
\label{EQN_fit}
\end{eqnarray}
with $x= 4\pi \zeta$. This gap appears to have eluded the previous analysis of \cite{kel17}. 

The low apparent dark matter content in \cite{gen17} arises from clustering of $\zeta$ close to the transition point $\zeta=1$,
that agrees with SPARC but deviates from MUGS2. Conversely, there is no apparent
low dark matter content in high redshift galaxies of MUGS2.

The SPARC-MUGS2 gap is expected to be generic for CDM galaxy models in $\Lambda$CDM, resulting from smoothness
inherent to diffusion by small angle gravitational scattering. At $6\sigma$, this discrepancy appears to be fundamental to the nature of 
CDM, unless perhaps the mass of the putative dark matter particle is anomalously small. The apparent $C^0$ galaxy dynamics 
in SPARC, however, points to a departure of Newton's second law as inertia drops at accelerations below $a_{dS}$,
when inertial mass-energy $U$ reduces to gravitational binding energy to the cosmological horizon ${\cal H}$. 

While a reduced inertia obviates the need for CDM in galaxies, a cosmological distribution of CDM is still required in light of 
the three-flat condition $\Omega_{M}+\Omega_\Lambda=1$ on the dimensionless densities of dark matter $(\Omega_M)$ 
and dark energy $(\Omega_\Lambda)$. The Compton wave length of the putative dark matter particle, greater 
than the scale of galaxies, may reach the scale of galaxy clusters.

In light of the above, we anticipate that the apparent self-similarity and $C^0$ galaxy dynamics 
shown in Figs. 2-3 extends to elliptical galaxies, which may be obtained through future studies
given the very large samples of elliptical galaxies available from, e.g., the 
{\em Sloan Digital Sky Survey} \citep{abo18}.

{\em Acknowledgements.} The author thanks B. Keller for kindly providing the MUGS2 data shown in Fig. 2,
the anonymous reviewer for detailed comments which greatly improved this manuscript, and K.-H. Chae for 
stimulating discussions on interpolation functions.

This research is supported in part by the National Research Foundation of Korea 
     (No. 2015R1D1A1A01059793, 2016R1A5A1013277 and 2018044640)
     
%\clearpage

\end{document}